%% file: icc_main.tex
\documentclass[sigconf,authorversion,nonacm,natbib=false]{acmart}
\usepackage{subfigure}
\usepackage{acronym}
\usepackage{comment}
\usepackage{makecell}

\usepackage[style=ieee,backend=bibtex,,maxbibnames=10]{biblatex}
\bibliography{biblio}

\begin{document}

\title{Insights from the Design Space Exploration of Flow-Guided Nanoscale Localization}


\author{Filip Lemic, Gerard Calvo Bartra, \\ Xavier Costa Pérez\footnotemark\authornote{X. Costa is also with NEC Laboratories Europe GmbH, Germany and ICREA, Spain.}}
\affiliation{%
  \institution{AI-Driven Systems Lab, i2Cat Foundation, Spain}
  \country{}}
\email{{name.surname}@i2cat.net}

\author{Arnau Brosa López, Sergi Abadal}
\affiliation{%
  \institution{NaNoNetworking Center in Catalunya, Polytechnic University of Catalonia, Spain}
  \country{}}
\email{{surname}@ac.upc.edu}

\author{Jorge Torres Gomez, Falko Dressler}
\affiliation{%
  \institution{TU Berlin, Germany}
  \country{}}
\email{{name.surname}@tu-berlin.de}

\author{Jakob Struye}
\affiliation{%
  \institution{University of Antwerp - imec, Belgium}
  \country{}}
\email{jakob.struye@uantwerpen.be}

\renewcommand{\authors}{F. Lemic, G. Calvo Bartra, A. Brosa López, J. Torres Gómez, J. Struye, F. Dressler, S. Abadal, X. Costa Pérez}

\renewcommand{\shortauthors}{Lemic et al.}

\begin{abstract}
Nanodevices with Terahertz (THz)-based wireless communication capabilities are providing a primer for flow-guided localization within the human bloodstreams. Such localization is allowing for assigning the locations of sensed events with the events themselves, providing benefits along the lines of early and precise diagnostics, and reduced costs and invasiveness. Flow-guided localization is still in a rudimentary phase, with only a handful of works targeting the problem. Nonetheless, the performance assessments of the proposed solutions are already carried out in a non-standardized way, usually along a single performance metric, and ignoring various aspects that are relevant at such a scale (e.g., nanodevices’ limited energy) and for such a challenging environment (e.g., extreme attenuation of in-body THz propagation). As such, these assessments feature low levels of realism and cannot be compared in an objective way. Toward addressing this issue, we account for the environmental and scale-related peculiarities of the scenario and assess the performance of two state-of-the-art flow-guided localization approaches along a set of heterogeneous performance metrics such as the accuracy and reliability of localization.
\end{abstract}

\ccsdesc[500]{Networks~Network architectures}
\ccsdesc[500]{Networks~Network simulations}
\ccsdesc[500]{Networks~Location based services}
\ccsdesc[300]{Human-centered computing}


\maketitle

\input{acronym_def} 
\vspace{-1mm}
\input{introduction}

\vspace{-1mm}
\input{flow_guided_loc}
\input{evaluation_setup}

\input{results}

\vspace{-1mm}
\input{conclusion}

\vspace{-1mm}
\section*{Acknowledgments}

This work was supported in part by the project IoBNT, funded by BMBF under grant number 16KIS1986K, and by the project NaBoCom, funded by DFG under grant number DR 639/21-2.

\renewcommand{\bibfont}{\footnotesize}
\printbibliography

\end{document}

%% file: acronym_def.tex
\acrodef{ML}{Machine Learning}
\acrodef{THz}{Terahertz}
\acrodef{GNN}{Graph Neural Networks}
\acrodef{ZnO}{Zinc Oxide}
\acrodef{IMU}{Inertial Measurement Unit}
\acrodef{RF}{Radio Frequency}
\acrodef{SINR}{Signal to Interference and Noise Ratio}
\acrodef{NN}{Neural Network}

%% file: introduction.tex

\section{Introduction}

Advancements in nanotechnology have paved the way for nanoscale devices that integrate sensing, computing, and data and energy storage capabilities~\cite{jornet2012joint}. 
These nanodevices are expected to enable various applications in precision medicine ~\cite{abbasi2016nano}. 
Some of these applications involve deploying nanodevices in patients' bloodstreams, which necessitates the nanodevice's size to be comparable to that of red blood cells (i.e., smaller than 5 microns). 
Due to their small physical size, these nanodevices can only rely on scavenging environmental energy, such as from heartbeats or ultrasound-based power transfer, using nanoscale energy-harvesting entities like \ac{ZnO} nanowires~\cite{jornet2012joint}. Consequently, these devices are expected to be passively flowing within the bloodstreams.

Recent advancements in advanced materials, particularly graphene and its derivatives~\cite{abadal2015time}, have opened up possibilities for nanoscale wireless communication in the \acf{THz} frequencies (i.e., 0.1-10~THz)~\cite{lemic2021survey}. 
Wireless communication capabilities enable two-way communication between nanodevices and the external world~\cite{dressler2015connecting}. 
Integrated nanodevices with communication capabilities are enabling sensing-based applications like oxygen sensing in the bloodstream for hypoxia detection (a biomarker for cancer diagnosis), as well as actuation-based applications like non-invasive targeted drug delivery for cancer treatment.
Nanodevices with communication capabilities also provide a primer for flow-guided localization within the bloodstream~\cite{lemic2021survey}. 
Flow-guided localization would allow associating the location of an event detected by a nanodevice, offering benefits along the lines of non-invasiveness, early and precise diagnostics, and reduced costs~\cite{gomez2022nanosensor,simonjan2021body,lemic2022toward}.

Performance evaluations of existing flow-guided localization approaches, specifically those described in~\cite{gomez2022nanosensor} and~\cite{simonjan2021body}, have been conducted in a simplified manner, focusing primarily on the mobility of nanodevices.
Consequently, these assessments overlook various potential effects of \ac{THz} wireless communication, such as interference, as well as energy-related constraints arising from energy-harvesting and intermittent operation of the nanodevices. 
It is also worth noting that~\cite{lemic2022toward} performed a limited performance evaluation, examining the number of nanodevices required for localizing a nanodevice at any location in the body through multi-hopping. 
As such, current evaluations provide only rough indications due to their limited realism and subjective evaluation methodologies.

In this article, we aim to enhance such assessments' realism by simultaneously considering multiple factors. 
These factors include accounting for the nanodevices' mobility, in-body nanoscale THz communication between them and the external world, and various energy-related and technological constraints, such as pulse-based modulation, which impact the nanodevices' performance. 
By incorporating these elements through the utilization of a simulator for objective and standardized performance evaluation of flow-guided nanoscale localization~\cite{lopez2023toward}, we seek to provide a more comprehensive and realistic understanding of the performance of flow-guided localization.
To the best of our knowledge, this article represents one of the first attempts at an objective and realistic performance assessment of different flow-guided localization approaches across a set of standardized heterogeneous metrics, including point and region estimation accuracies and localization reliability.
Our work is inspired by established approaches targeting objective evaluation and benchmarking of traditional indoor localization solutions stemming from efforts such as NIST PerfLoc~\cite{moayeri2016perfloc}, EU EVARILOS~\cite{van2015platform}, and Microsoft/IPSN indoor localization competition~\cite{lymberopoulos2015realistic}.

%% file: flow_guided_loc.tex

\section{Flow-guided Localization}

\subsection{Flow-guided Localization Fundamentals}

Flow-guided localization aims to localize a target event using nanodevices without requiring the nanodevices to determine their own location.
The concept presented in~\cite{lemic2022toward} supports this type of scenario and falls within this category. 
However, the notable representatives of this localization approach are~\cite{gomez2022nanosensor} and~\cite{simonjan2021body}. 
In these studies, \ac{ML} models are employed to differentiate the regions through which each nanodevice passes during a single circulation through the bloodstream. 
The authors in~\cite{simonjan2021body} achieve this by tracking the distances a nanodevice covers during its circulation using a nanoscale \ac{IMU}. 
However, this poses challenges concerning the limited resources available for storing and processing \ac{IMU}-generated data at the nanodevice level, as well as the accuracy of IMU readings being affected by the blood's vortex flow. 
On the other hand, the authors in~\cite{gomez2022nanosensor} address these challenges by tracking the time required for each circulation through the bloodstream. 
The captured distance or time information is then transmitted to a beaconing anchor located near the heart using short-range \ac{THz}-based backscattering.

Unlike~\cite {lemic2022toward}, these localization approaches are not explicitly designed to provide precise point localization of the target. 
Despite the potential benefits of achieving point localization for healthcare diagnostics, these methods focus on detecting the body region through which the nanodevice has passed. 
Furthermore, increasing the number of circulations the nanodevices make through the bloodstream can improve the accuracy and reliability of region detection. 
However, this increase would result in higher energy consumption for the localization procedure. 
Thus, performance metrics such as point and region accuracies and reliability should and will be evaluated in relation to the application-specific delay allowed for localizing events, as also stated in~\cite{lopez2023toward}.

\subsection{Off-the-shelf Localization Solution} 

The first flow-guided localization solution we consider is an off-the-shelf solution presented in~\cite{gomez2022nanosensor}.
The authors assume that, due to the physiology of the bloodstream, the nanodevices travel in closed loops. 
They further assume that the nanodevices feature an internal counter that increases its value periodically and restarts in each passage through the heart. 
The final counter value is reported to an external gateway located at the heart proximity before resetting the counter.
The counter value is used as an input to a \ac{ML} model to predict the traveling loop for each nanodevice. 
The utilized \ac{ML} model is a \ac{NN} that implements two fully connected layers, a ReLU activation function for the first and a SoftMax function for the second layer. 
The model is trained to classify 24 different circuits with traveling times as reported by the simulator (more details in the following section). 
The optimization algorithm to evaluate the \ac{NN} hyperparameters is the Limited Memory Broyden–Fletcher–Goldfarb–Shanno (L-BFGS), which minimizes the cross-entropy between predictions and labels.

\subsection{In-house Localization Solution} 

The second considered approach is a a scenario-optimized in-house \ac{NN} solution that implements three fully connected layers, with PReLU activation function for the first two and log-softmax for the last one. 
The first and second layers feature a dropout for regularization and batch normalization to stabilize the learning process. 
The hidden layer's size is 512, and the model is trained to classify 25 classes (in contrast to the off-the-shelf solution, which is, by design, unable to detect events in the heart). 
We use the Negative Log Likelihood loss due to its ability to handle unbalanced datasets. 
Finally, we use the Adam optimizer, as it adjusts the learning rate dynamically and is known to operate well with relatively simple fine-tuning of the hyperparameters.

\begin{figure*}
\centering
\includegraphics[width=0.72\linewidth]{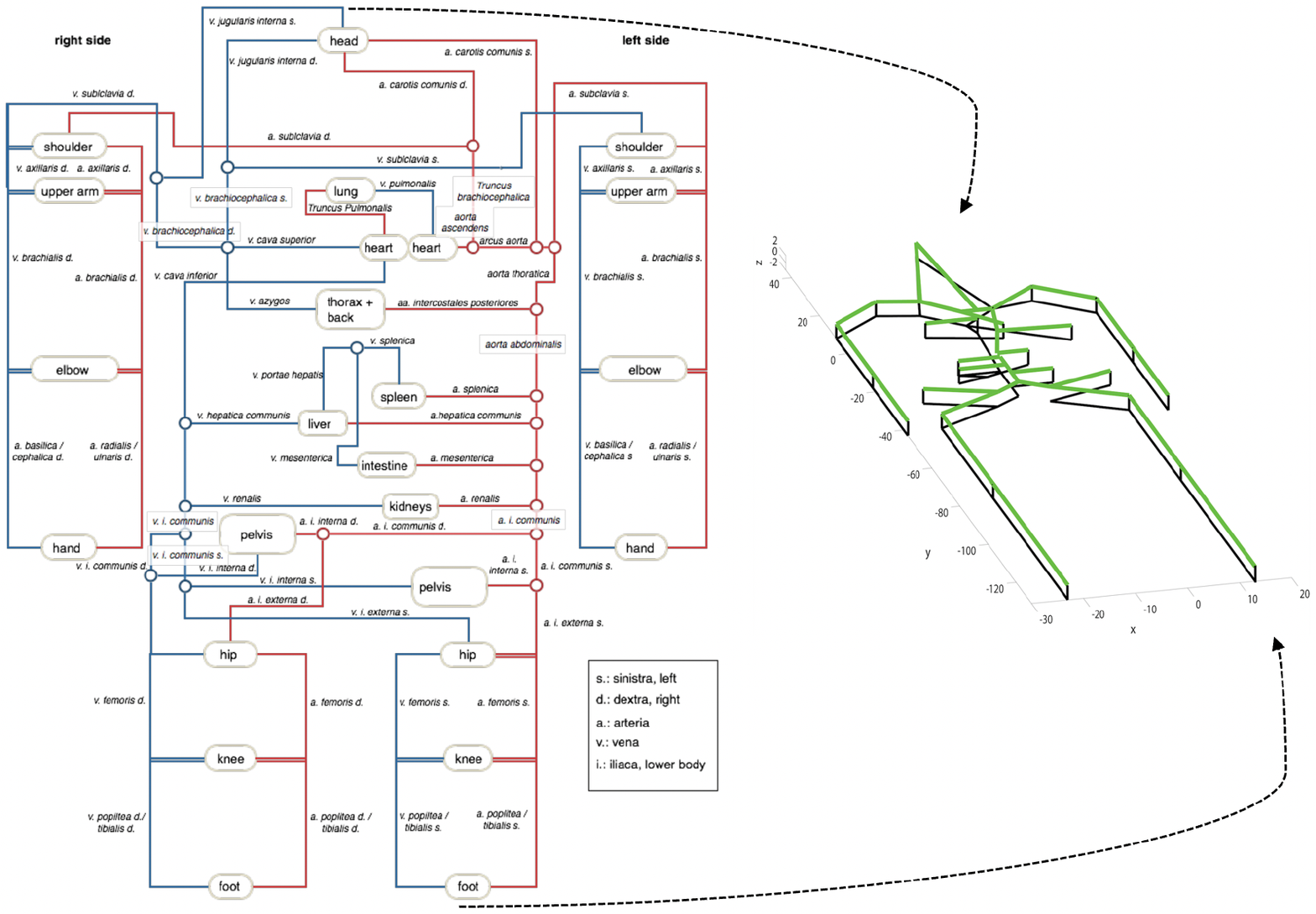}
\caption{Nanodevice mobility in the BloodVoyagerS~\cite{geyer2018bloodvoyagers}}
\label{fig:bloodvoyagers}
\end{figure*}    

%% file: evaluation_setup.tex
\section{Evaluation Framework}
\label{sec:analysis}

In this section, we outline the evaluation framework employed to assess the effectiveness of two flow-guided localization approaches. The framework integrates a carefully designed simulator, specific scenarios and parameters defining our design space, and a set of performance metrics to rigorously test the capabilities of these approaches under simulated physiological conditions.

Our framework enables realistic testing of nanodevice functionality and localization techniques, incorporating wireless communication and energy harvesting. By simulating both the physiological and mechanical properties of the human vascular system, this setup addresses the challenges nanodevices face in real scenarios, including modeling THz signal propagation and energy dynamics within the bloodstream, essential for deploying such technologies in medical diagnostics and treatments.

The framework is implemented in the form of a simulator~\cite{lopez2023toward}. 
The simulator assumes that deploying a flow-guided localization in the bloodstream requires at least one anchor attached to the patient's body. 
Approaches in~\cite{gomez2022nanosensor} and~\cite{simonjan2021body} can be enabled with a single anchor positioned near the heart, as the heart is the only body region with guarantees that each nanodevice will pass through it in every iteration. The anchor is envisioned as a static entity powered by reliable sources such as batteries; hence, it is assumed to be continuously operational. 
The role of the anchor is to transmit beacons and receive backscattered responses from the nanodevices.

The nanodevices are energy-harvesting entities that can move passively in the bloodstream. 
To model their mobility, the utilized simulator features the integration with BloodVoyagerS~\cite{geyer2018bloodvoyagers}.
BloodVoyagerS provides a simplified model of the bloodstream, including 94 vessels and organs, with the coordinate system centered in the heart. 
All organs have the same spatial depth, with a reference thickness of 4~cm resembling the depth of a kidney. 
This results in the z-coordinates of the nanodevices ranging from 2 to -2~cm, as illustrated in Figure~\ref{fig:bloodvoyagers}. 
The simulator assumes that arteries and veins are positioned anterior and posterior, respectively. 
Transitions between arteries and veins occur in the organs, limbs, and head. 
In the heart, blood transitions from veins to arteries (i.e., posterior to anterior). 
The flow rate is modeled based on the relationship between pressure difference and flow resistance, resulting in average blood speeds of 20~cm/sec in the aorta, 10~cm/sec in arteries, and 2-4~cm/sec in veins. 
Transitions between arteries and veins are simplified by assuming a constant velocity of 1~cm/sec.

In the simulator, the nanodevices are assumed to have capacitors for energy storage and utilize \ac{ZnO} nanowires for energy harvesting. 
The charging of capacitors is modeled as an exponential process that considers the energy-harvesting rate and interval (e.g., 6 pJ per sec and per 20 ms for harvesting from heartbeats and ultrasound-based power transfer, respectively~\cite{jornet2012joint}), as well as the storage capacity of the capacitors. 
The nanodevices exhibit intermittent behavior due to energy harvesting and storage constraints. 
This behavior is modeled through a \emph{Turn ON} threshold, where a nanodevice turns on if its current energy level exceeds the threshold. 
Once the energy is fully depleted, the nanodevice turns off and turns back on when its energy increases above the threshold.

When the nanodevices are turned on, they periodically perform sensing or actuation tasks at a given sampling frequency or granularity. 
Each execution consumes a constant amount of energy, meaning that more frequent tasks result in higher energy consumption. 
The location of the event to be detected is assumed to be hard-coded by the experimenter. 
A nanodevice is considered to detect an event if it is turned on and the Euclidean distance between its location at the time of task execution and the location of the event is smaller than a predefined detection threshold.

In communication with a nanodevice, the anchor transmits beacons at a constant frequency and power.  
The nanodevices passively receive the beacons and actively backscatter a response, which consumes energy. 
The backscattered packets from the nanodevices contain two pieces of information: the time elapsed since their last passage through the heart and an event bit~\cite{pascual2024math}. 
These raw data points are then used by a flow-guided localization approach to localize an event. 
Whenever a nanodevice passes through the heart, the time elapsed since the last passage is reset to avoid accumulating multiple circulation periods. 
The event bit is set to logical "1" if an event is successfully detected, and is envisaged to be reset in each passage through the heart.

The \ac{THz} channel is modeled by calculating the receive power for each pair of communicating devices and scheduling the invocation of the \emph{ReceivePacket()} method based on the corresponding propagation time. 
The channel model takes into account in-body path-loss and Doppler effects. 
The path-loss is determined by considering the attenuation and thickness parameters of the vessels, tissues, and skin. 
The Doppler effect is incorporated by evaluating the changes in relative positions between the nanodevices and the anchor over time. 
The \emph{ReceivePacket()} method checks for potential collisions by calculating the \ac{SINR} and discards the packet if the SINR falls below a predefined threshold for reception, known as the receiver sensitivity.

Our aim is to explore several aspects of the design space of flow-guided localization.
To this end, we identify a set of parameters worth exploring and keep the others at a fixed value. Table~\ref{tab:paramaters} outlines the fixed parameters. 
Then, the parameters constituting the considered design space are (i) the number of administered nanodevices, (ii) event sampling frequency or granularity, and (iii) the distance threshold for event detection. Table~\ref{tab:paramaters2} shows their baseline values, as well as values considered within the design space.
In each scenario, we vary one parameter while keeping the others at their baseline values, allowing us to isolate the effects of a single parameter on the overall performance of the considered solutions. 

Following guidelines from~\cite{lopez2023toward}, we utilize three heterogeneous metrics for characterizing the performance of the evaluated solutions, which provide insights into different performance trade-offs that occur in flow-guided localization.
The first one is the \textbf{region detection accuracy}, which characterizes the level of correctness of estimating a body region containing an event. 
Then, \textbf{point accuracy} represents the Euclidean distance between the true location of an event and its estimated location as yielded by a flow-guided localization solution. 
Finally, the \textbf{reliability of localization} characterizes the capability of the solution to provide a location estimate (regardless of its correctness) after a certain period. 

To obtain the performance metrics, we first trained both localization solutions. The training was carried out by specifying the location of an event in the centroid of each of the 25 body regions covered by BloodVoyagerS (i.e., organs, head, and extremities, with regions indicated with gray rounded rectangles in Figure~\ref{fig:bloodvoyagers}). 
The raw training data for each event location was then generated by running the simulations for 5000~sec, and fed in both of the considered solutions.
This was followed by generating a set of testing datasets to capture the outlined design space.
The testing dataset for each scenario contained 25 different event locations, one in each region, selected randomly within the region. 
The intuition for such a selection is that an event of interest can be located in any part of the considered region.
Given that the considered solutions are unable to provide a point estimate but solely an estimated region by design, the point estimate was obtained as the centroid location of the estimated region.

\begin{table}
\small
\begin{center}
\caption{Simulation parameters}
\vspace{-1mm}
\label{tab:paramaters}
\begin{tabular}{l r}
\hline
\textbf{Parameter} & \textbf{Value} \\
\hline
Generator voltage $V_g$ [V] & 0.42 \\
Energy consumed in pulse reception $E_{R_{X pulse}}$ [pJ] & 0.0 \\
Energy consumed in pulse transmission $E_{T_{X pulse}}$ [pJ] & 1.0 \\
Maximum energy storage capacity [pJ] & 800 \\
Turn ON/OFF thresholds [pJ] & 10/0 \\
Harvesting cycle duration [ms] & 20 \\
Harvested charge per cycle [pC] & 6 \\
Transmit power $P_{T_X}$ [dBm] & -20 \\
Operational bandwidth [GHz] & 10 \\ 
Receiver sensitivity [dBm] & -110 \\ 
Operational frequency [THz] & 1 \\ 
Simulation time [sec] & 1100 \\
Number of anchors & 1 \\
\hline
\end{tabular}
\end{center}
\vspace{-2mm}
\end{table}

\begin{table}
\small
\begin{center}
\caption{Design space parameters}
\vspace{-1mm}
\label{tab:paramaters2}
\begin{tabular}{l p{1.2cm} r}
\hline
\textbf{Parameter} & \makecell{\textbf{Baseline}} & \textbf{Design space} \\
\hline
Number of nanodevices & \hfil 64 & [32, 64, 128] \\
Event sampling granularity [1/sec] & \hfil 3 & [2, 3, 5, 10] \\
Event detection threshold [cm] & \hfil 1 & [0.5, 1, 2, 3] \\
\hline
\end{tabular}
\end{center}
\vspace{-2mm}
\end{table}

Our results in Figures~\ref{fig:nr_nanobots},~\ref{fig:granularity} and~\ref{fig:threshold} are depicted in a way that each data point on the x-axis (e.g., each box-plot in the point accuracy plot) shows the performance averaged over the 25 evaluation points. 
Such depiction of the results allows for reporting the average performance of the solutions for the entire environment, encapsulating the performance variability in different environmental regions.

%% file: results.tex

\vspace{-1mm}
\section{Evaluation Results}
\label{results}

\begin{figure*}[!t]
\centering
\subfigure[Off-the-shelf solution]{
\includegraphics[width=0.44\textwidth]{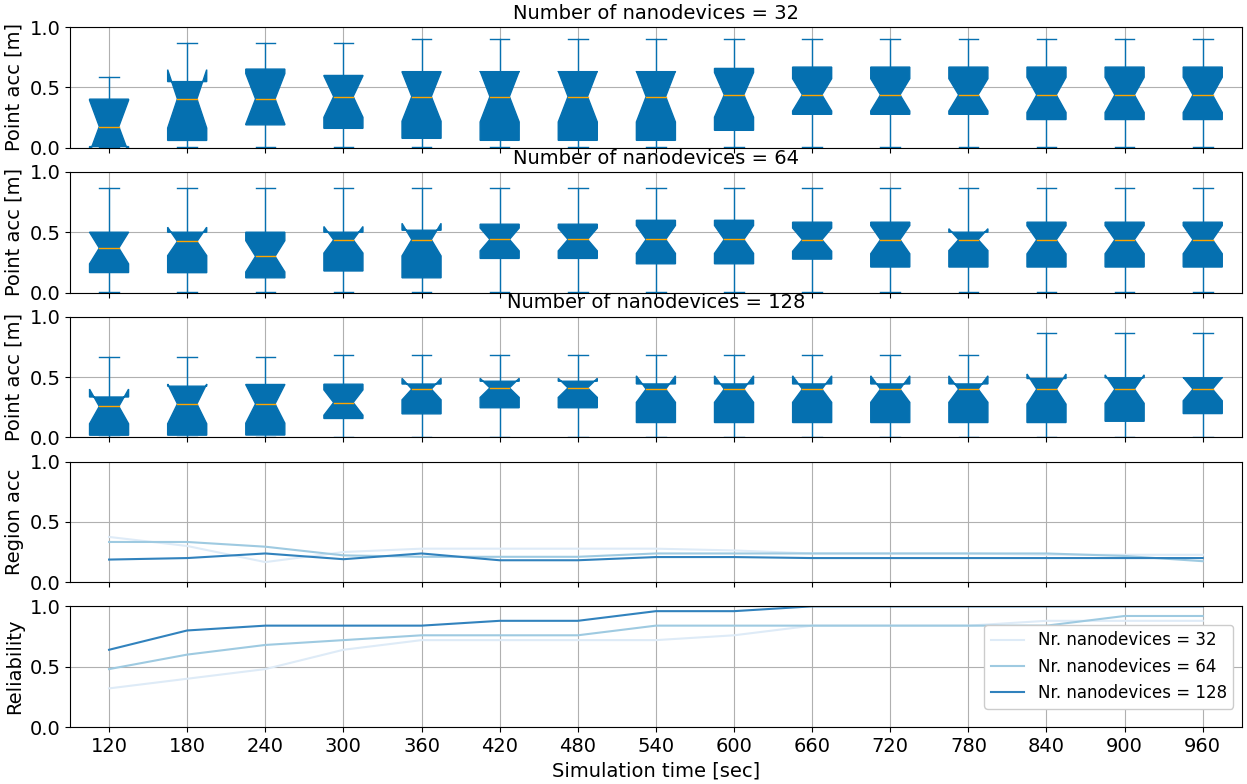}}
\subfigure[In-house solution]{
\includegraphics[width=0.44\textwidth]{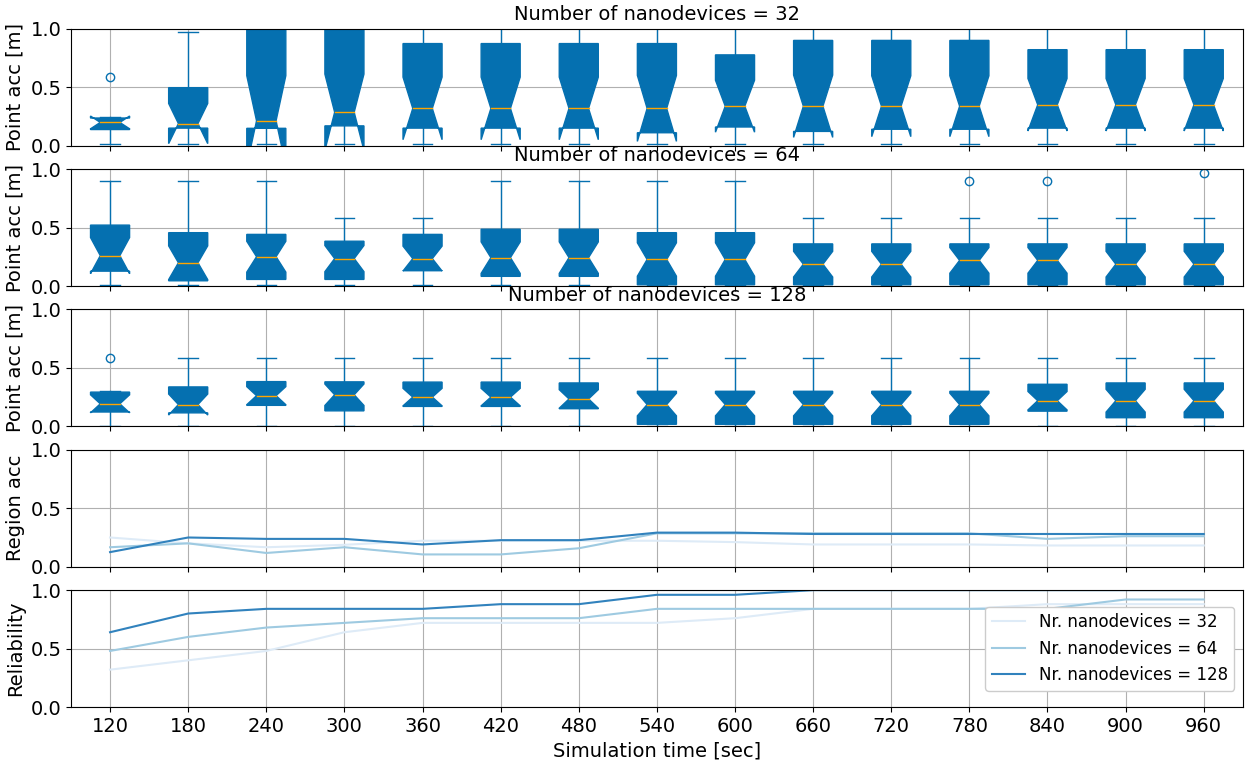}}
\vspace{-3mm}
\caption{Number of nanodevices}
\label{fig:nr_nanobots}
\vspace{-2mm}
\end{figure*}

\begin{figure*}[!t]
\centering
\subfigure[Off-the-shelf solution]{
\includegraphics[width=0.44\textwidth]{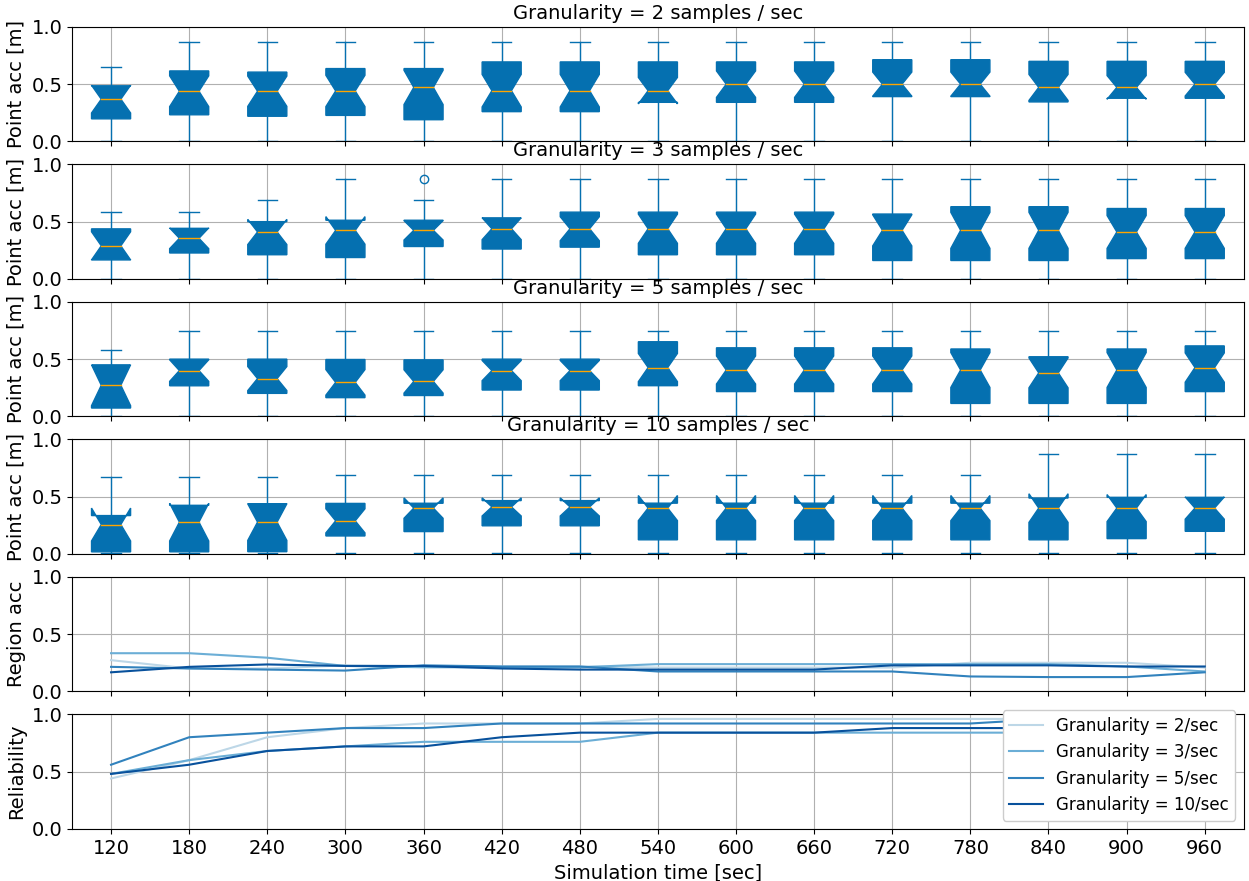}}
\subfigure[In-house solution]{
\includegraphics[width=0.44\textwidth]{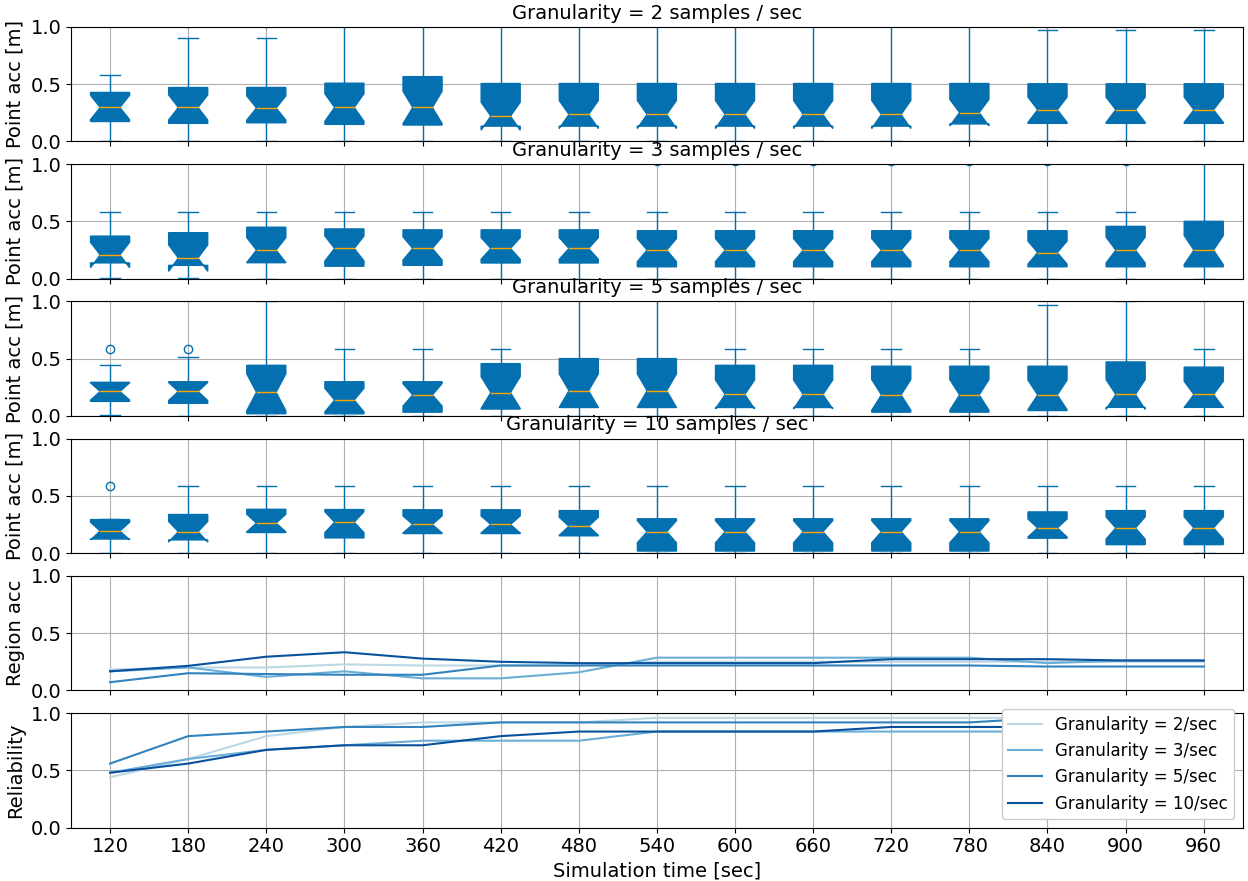}}
\vspace{-3mm}
\caption{Event sampling granularity}
\label{fig:granularity}
\vspace{-2mm}
\end{figure*}

\begin{figure*}[!t]
\centering
\subfigure[Off-the-shelf solution]{
\includegraphics[width=0.44\textwidth]{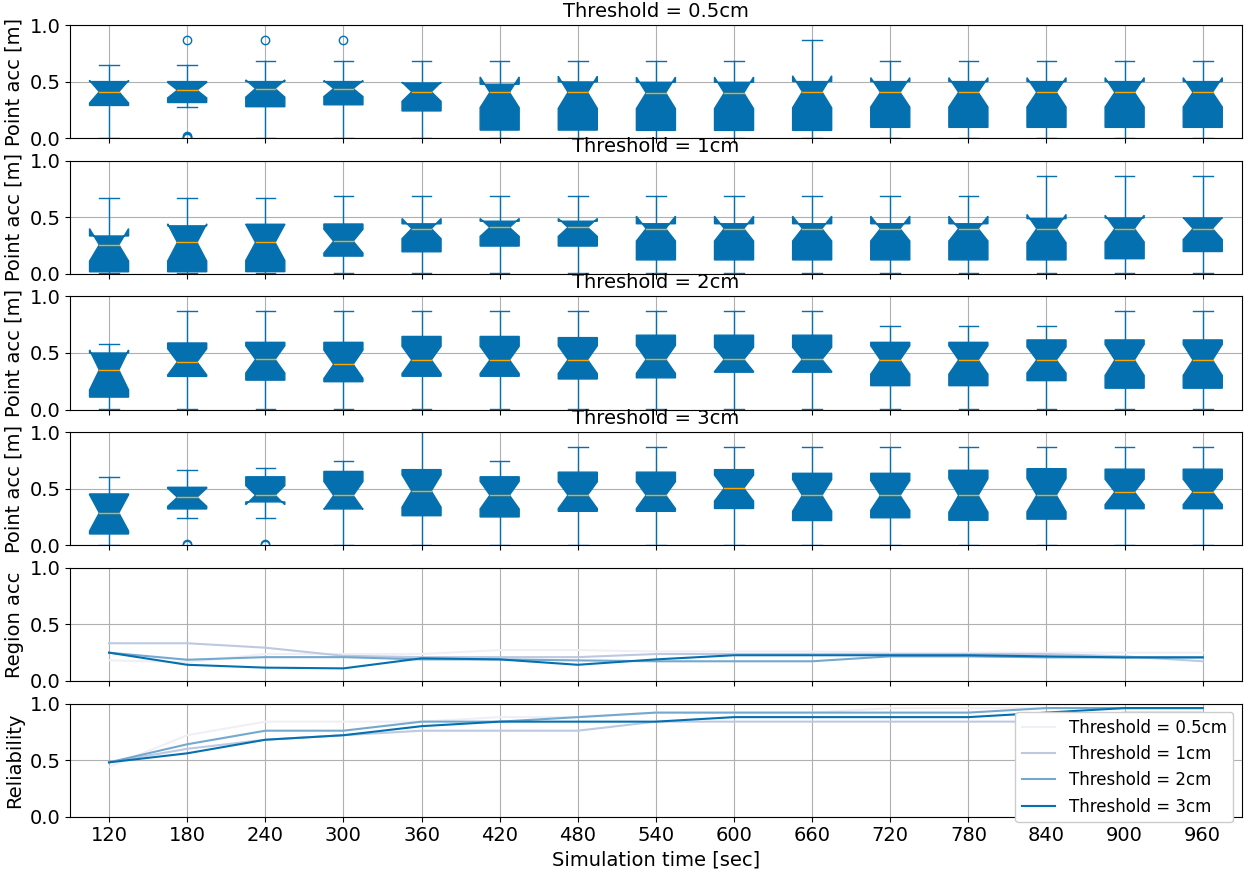}}
\subfigure[In-house solution]{
\includegraphics[width=0.44\textwidth]{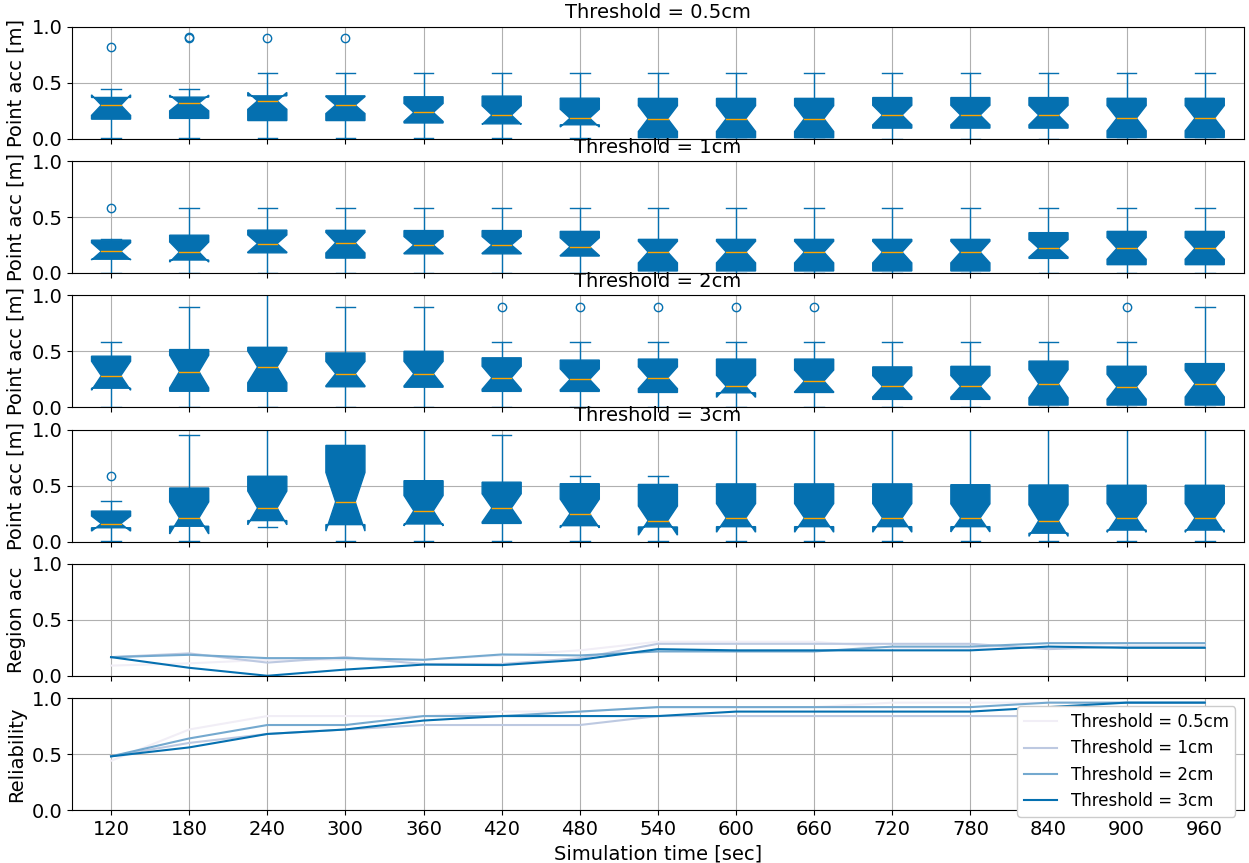}}
\vspace{-3mm}
\caption{Detection distance threshold}
\label{fig:threshold}
\vspace{-2mm}
\end{figure*}

Figure~\ref{fig:nr_nanobots} depicts the performance achieved by the two solutions as a function of the number of administered nanodevices.
The figure shows that the in-house solution generally outperforms the off-the-shelf solution regarding the point accuracy metric. 
For example, in the scenario assuming 128 deployed nanodevices, the in-house solution yields a median localization error of around 25~cm throughout the simulation run.
In contrast, this error is roughly 40~cm throughout the simulation for the off-the-shelf solution. 
This can be confirmed by the results depicted in Figures~\ref{fig:granularity} and~\ref{fig:threshold} demonstrating the influence of the event sampling granularity and event detection threshold of the performance of the considered solutions, respectively.
A similar conclusion can be drawn when observing the region detection accuracy, although in this case, the improvements yielded by the in-house solution are less pronounced and in the range of several percent. 
This is observed because the off-the-shelf solution is not tailored to our framework data. As such, its hyperparameters have not been fine-tuned for the raw datasets utilized in this study, in contrast to the in-house-made solution.

Moreover, Figure~\ref{fig:nr_nanobots} depicts that point and region estimation accuracies increase either slightly or not at all with the increase in the duration of a simulation runtime. 
For example, considering the in-house solution with the assumption of 64 nanodevices being deployed in the bloodstream, the median localization error characterizing the pointing accuracy decreases from around 30 to 20 cm if the duration of raw data collection is extended from 120 to 960 sec.
At the same time, the region estimation accuracy increases from around 20 to 25\%.
Interestingly, for the scenario assuming 128 deployed nanodevices, the point and region detection accuracies are effectively unchanged throughout the simulation run.
The main reason for such behavior stems from two aspects affecting the solutions' performance.
The first and arguably more intuitive reason comes from the fact that \ac{ML} models generally improve their performance with the increased amount of raw data fed into them for making predictions.
Obviously, an increase in the duration of a simulation run results in an increase in the amount of raw input data for the considered solutions, which benefits estimation accuracy.
However, one should also account for the fact that \ac{THz} communication between the nanonodes and the anchor is challenging for several reasons, including high in-body attenuation, high mobility of the nanonodes, and self-interference between different nanonodes trying to communicate with the anchor simultaneously. 
For these reasons, the communication is unreliable, which can result in the anchor not receiving the raw data by some nanonodes at specific time instances.
More problematically, these nanonodes do not reset their iteration times and event bits in such cases.
Once the data is eventually reported to the anchor, the reported iteration times are the compounds of multiple iterations, while the event bit might be erroneous (i.e., the event was detected in one of the iterations, yet propagated through multiple iterations, some of which did not feature the event).
In addition, the nanodevice's energy-harvesting nature and, consequently, its intermittent operation might result in the nanodevice failing to detect the event as it was turned off, despite passing through the bloodstream region containing such an event.
More details on the features of the raw data can be found in~\cite{lopez2023toward}.
The outlined behavior results in the fact that, although more data inputted into the models should increase the accuracy of estimation, its highly erroneous nature balances out these improvements, resulting in the ``flat'' performance regarding region and point detection accuracy for both solutions.

Observing Figures~\ref{fig:nr_nanobots},~\ref{fig:granularity}, and~\ref{fig:threshold}, one can see that the region estimation accuracy respectively converges towards the values of roughly 20 and 25\% for both localization solutions, although the rate of convergence differs across scenarios. 
The converging values represent the achievable performance of the models, which is in contrast to the significantly better performance reported by state-of-the-art works proposing the considered models (e.g.,~\cite{gomez2022nanosensor}).
The models' unsatisfactory accuracy is due to a difference in the raw data utilized for their training and evaluation. The raw input data generated in these works did not account for the energy-harvesting nature of the nanodevices, as well as for unreliable communication between the nanodevices and the anchor resulting from various phenomena such as self-interference between the nanodevices, high-attenuation of in-body \ac{THz} communication, etc.
In addition, these models are, by design, unable to distinguish between the left and right-hand sides of the body, given that in such cases, the iteration times are the same.
Future efforts should consider new and more powerful ML tools, with primary candidates stemming from the \ac{GNN} family due to their intrinsic ability to handle complex graph-structured data, capture non-linear relationships between different data points (i.e., nodes), and generalize to unseen data. 
In addition, future efforts should consider introducing additional anchors at strategic locations of the body (primarily the hands and leg wrists) to support the ascertainment of the left and right-hand sides of different body regions.

Regarding the design space of the considered solutions, Figure~\ref{fig:nr_nanobots} depicts the effects of an increasing number of administered nanodevices on the accuracy and reliability-related metrics.
As visible, an increase in the number of nanodevices increases the accuracies of both solutions for point and region estimation.
As an example, when increasing the number of nanodevices from 32 to 128 and considering an entire simulation run with the duration of 960~sec, the third quartile of localization error for the in-house solution is decreased from more than 75 to less than 40~cm.
This improvement is a direct result of the system producing more raw data as a result of introducing additional nanodevices.
However, the region accuracy is only minorly affected, which can be attributed to the fact that the more data is generated, the more erroneous instances it features, eventually resulting in poor region detection accuracy.
The introduction of additional nanodevices benefits reliability, although eventually, the reliability of producing location estimates generally converges toward 100\%.    
Optimizing the reliability increase will benefit applications with latency requirements, e.g., drug delivery.

Figures~\ref{fig:granularity} and~\ref{fig:threshold} depict the effects of the event sampling frequency and detection distance threshold on the performance of the solutions, respectively.
As visible, these parameters' variations only slightly affect the considered metrics. 
The sampling granularity only affects the rate at which the events can be detected. 
As such, it is expected that the more frequent the sampling, the higher the convergence rate toward the region detection accuracy achievable by the models.
In addition, the energy consumed in sampling is small compared to the overall energy of the nanodevices.
In other words, even the fastest sampling of 10 samples/sec is insufficient to deplete the energy of the nanodevices; hence, the granularity does not significantly affect the reliability of producing estimates. 
More generally, the selection of sampling frequency should be considered in relation to the blood speeds and available energy at the nanonode level.
For the considered regions in which the blood speeds are 1~cm/sec and with the event detection threshold of 1~cm, there is no need for increasing the sampling frequency beyond 1 sample/sec, as this will solely result in increased energy consumption.

As visible in Figure~\ref{fig:threshold}, an increase in the distance threshold for event detection harms the point accuracy. When the detection distance is increased, events that are further away from the nanodevices are also detected. This has a direct impact on the accuracy of event detection.
For certain regions, such as the ones in the torso, this increase is sufficient to detect events in regions where a nanodevice did not pass, eventually resulting in some estimates featuring significant errors.  
Thus, the distribution of errors depicted in each point accuracy box-plot gets visibly larger when the threshold is increased beyond 1~cm.

%% file: conclusion.tex

\vspace{-1mm}
\section{Conclusion}

We have followed a standardized methodology for assessing the performance of two contemporary flow-guided localization solutions.
The followed assessment methodology allowed us to explore the design space of the solutions in an objective way and along heterogeneous metrics.
The high level of realism was achieved by utilizing a state-of-the-art simulation environment that is able to capture the peculiarities of nanodevices, \acf{THz} wireless nanocommunication, and the harsh environment that the human bloodstream represents.
Our results reveal relatively poor performance of the solutions, which can be attributed to the unreliable nature of \ac{THz} communication between the in-body nanodevices and the outside world, and the inability of the solutions to deal with the high complexity and erroneous nature of the input data.
Future work will aim at enhancing the accuracy of localization by considering different \ac{ML} models, primarily the ones from the \acf{GNN} family, and introducing additional anchors at strategic locations on the body.